\documentclass[graybox]{svmult}

\input{tcilatex}
\newcommand{\prd}{Phys.~Rev.~D}
\newcommand{\prl}{Phys.~Rev.~Lett.}

\usepackage{mathptmx}       
\usepackage{helvet}         
\usepackage{courier}        
\usepackage{type1cm}        
\usepackage{makeidx}         
\usepackage{graphicx}        
\usepackage{multicol}        
\usepackage[bottom]{footmisc}

\makeindex             

\begin{document}

\title*{Thick Dirac-Nambu-Goto branes on black hole backgrounds}
\author{Viktor G. Czinner}
\institute{Viktor G. Czinner$^{1,2}$ \at $^1$Centro de Matem\'atica, Universidade do Minho, 
Campus de Gualtar, 4710-057 Braga, Portugal;\\
$^2$HAS Wigner Research Centre for Physics, Institute for Particle and Nuclear Physics,
H-1525 Budapest, P.O.Box 49, Hungary.\\
\email{czinner.viktor@wigner.mta.hu}}

\maketitle

\abstract{Thickness corrections to static, axisymmetric Dirac-Nambu-Goto branes 
embedded into spherically symmetric black hole spacetimes with arbitrary number 
of dimensions are studied. First, by applying a perturbative approximation, it is found 
that the thick solutions deviate significantly in their analytic properties from the 
thin ones near the axis of the system, and perturbative approaches around the thin 
configurations can not provide regular thick solutions above a certain dimension. 
For the general case, a non-perturbative, numerical approach is applied and regular 
solutions are obtained for arbitrary brane and bulk dimensions. As a special case, it 
has been found that 2-dimensional branes are exceptional, as they share their analytic 
properties with the thin branes rather than the thick solutions of all other dimensions.}

\section{Introduction}\label{int}
The study of higher dimensional black holes, branes and their interactions is an active 
field of research in several different areas of modern theoretical physics. One interesting 
direction, which has been first introduced by Frolov \cite{Frolov}, is to consider a brane - 
black hole toy model for studying merger and topology changing transitions in higher 
dimensional classical general relativity \cite{Kol}, or in certain strongly coupled 
gauge theories \cite{MMT} through the gauge/gravity correspondence. 
The model consists of a bulk $N$-dimensional 
black hole and a test $D$-dimensional brane in it ($D\leq N-1$), called {\it brane-black 
hole} (BBH) system. The brane is infinitely thin, and it is described by the Dirac-Nambu-Goto 
\cite{DNG} action. 

\begin{figure}
\sidecaption
\includegraphics[scale=.89]{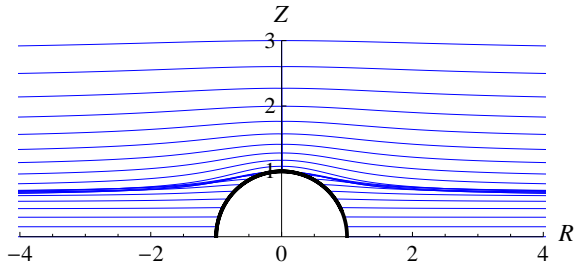}
\caption{A sequence of thin brane equilibrium configurations on Schwarzschild
background. The different configurations belong to different boundary conditions. 
For simplicity, the horizon radius is put to be 1, and $R$ and $Z$ 
are standard cylindrical coordinates.}
\label{fig1}   
\end{figure}

Due to the gravitational attraction of the black hole, 
the brane is deformed and there are two types of equilibrium configurations. The brane either 
crosses the black hole horizon (supercritical), or it lies totally outside of the black hole 
(subcritical), see Fig.~\ref{fig1}. 

Generalizations of the BBH model by studying the effects of thickness corrections obtained from 
higher order curvature terms in the effective brane action, have also been studied by Frolov 
and Gorbonos \cite{FG} within a perturbative approach, near the critical solution and restricted to 
the Rindler zone. As an interesting result they found that when the spatial dimension of the brane is 
greater than 2, supercritical solutions behave quite differently from subcritical ones, and according to 
their numerical analysis, they did not find evidence for the existence of such solutions. They suggested 
that quantum corrections may cure this pathological behaviour.

As a different approach, in three consecutive papers \cite{CF, Cz1, Cz2}, we have re-considered the problem
of thickness corrections to the BBH system within a more general framework than that of \cite{FG}. We did 
not restrict 
ourselves to work neither in the Rindler zone, nor in the near-critical solution region, and we also chose to 
follow a different path in obtaining the brane Euler-Lagrange equation. As a result we were able to provide 
the complete set of regular solutions of the thick-BBH problem for arbitrary brane and bulk dimensions, and 
also to clarify the question of phase transition in the system. 

In this paper we present very briefly the outline of our main findings, and refer the reader to the works 
\cite{CF, Cz1, Cz2} for details and the complete results.


\section{Perturbative approach}\label{pert}

In \cite{CF}, we applied a linear perturbation method to the thick BBH problem obtained from the 
curvature corrected Dirac-Nambu-Goto action:
\begin{equation}
S=\int d^D\zeta\sqrt{-\mbox{det}\gamma_{\mu\nu}}\left[-\frac{8\mu^2}
{3 \ell}(1+C_1R+C_2K^2)\right]\ ,\nonumber
\end{equation}
and derived the general form of the perturbation equation for the thick branes. From the asymptotic 
behaviour of the 
perturbation equation, in accordance with the results of \cite{FG}, we found that 
there is no regular solution of the perturbed problem in the Minkowski embedding case, unless the 
brane is a string, or a 2-dimensional sheet. This restriction, however does not hold for the black 
hole embedding solutions, which are always regular within our perturbative approach. 

From these results we concluded that the absence of regular solutions above the dimension $D=3$ 
implies, that the problem can not be solved within perturbative approaches around the thin solutions 
which are \textit{not smooth} on the axis of the system. For a general discussion, one needs 
to find a new, non-perturbative solution of the problem, that is expected to behave differently 
from the thin solutions by being \textit{smooth} on the axis. 
 
\begin{figure}
\sidecaption[t]
\includegraphics[scale=.75]{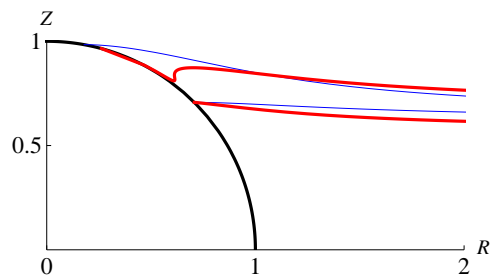}
\caption{The picture shows two thick (red) brane configurations together with their thin
(blue) counterparts in the case of an $N=5$, $D=4$ black hole embedding. The boundary 
conditions are $\theta_0=\tfrac{\pi}{4}$ (bottom curves) and $\tfrac{\pi}{17}$ (top curves).}
\label{fig2}
\end{figure}

After these conclusions, we provided the solution of the perturbation equation for various brane 
($D$) and bulk ($N$) dimensions. The far distance equations were integrated analytically, while the 
near horizon solutions were obtained by numerical computations. The deformations of the perturbed brane 
configurations were plotted (see eg.~Fig.~\ref{fig2}) and a comparison was made with the corresponding 
thin brane configurations with identical boundary conditions.

\section{Non-perturbative solutions}\label{nonpert}
In \cite{Cz1} we further studied the effects of curvature
corrections to the BBH system. Since the results of \cite{CF} clearly showed that 
perturbative approaches fail to provide regular solutions near the axis of the system 
in Minkowski type topologies, we considered a different, exact, numerical approach to 
the problem. We analysed the asymptotic properties of the complete, $4th$-order, 
highly non-linear Euler-Lagrange equation of the thick BBH system, presented its 
asymptotic solution for far distances, and obtained regularity conditions in the 
near horizon region for both Minkowski and black hole embeddings. We showed that the 
requirement of regularity for the thick solution 
defines almost completely the boundary conditions for the Euler-Lagrange equation in 
the Minkowski embedding case. The only exceptions are the brane configurations 
with 1, 2 and 3 spacelike dimensions. In the cases of 1 and 3, regular solutions of 
the problem could be found, however, which was an unexpected result, the problem could 
not be solved with the applied, non-perturbative method for the case of 2 spacelike 
dimensions.

\section{The 2-dimensional case}\label{2dim}
In \cite{Cz2} we further studied the problem of a topologically flat 2-brane in the thickness 
corrected BBH system. Despite the previously mentioned difficulties, we were able to find a 
regular, non-perturbative, numerical solution for this special case also, based on earlier 
perturbative considerations \cite{CF}. The main result here was the observation that the 
2-dimensional case is special as being \textit{non-analytic} at the axis of the system, 
just like the thin solutions. This property makes it unique in the family of thick solutions, 
as in all other dimensions both the Minkowski and black hole embedding solutions are 
\textit{analytic} in their entire domain. 


\begin{acknowledgement}
The research leading to these results was supported in part by the Japan Society for the 
Promotion of Science contract No.~P06816, the Hungarian National Research Fund, OTKA No.~K67790 
grant, and by the National Research Foundation of South Africa. It has also received funding from 
the European Union Seventh Framework Programme (FP7/2007-2013) under grant agreement n$^{\circ}$ 
PCOFUND-GA-2009-246542 and from the Foundation for Science and Technology of Portugal.
\end{acknowledgement}

\end{document}